\documentclass[useAMS]{mn2e}
\usepackage{psfig}

\usepackage{times}

\def\spose#1{\hbox to 0pt{#1\hss}}
\newcommand\lsim{\mathrel{\spose{\lower 3pt\hbox{$\mathchar"218$}}
     \raise 2.0pt\hbox{$\mathchar"13C$}}}
\newcommand\gsim{\mathrel{\spose{\lower 3pt\hbox{$\mathchar"218$}}
     \raise 2.0pt\hbox{$\mathchar"13E$}}}
\def\ltsima{$\; \buildrel < \over \sim \;$}
\def\lsim{\lower.5ex\hbox{\ltsima}}
\def\gtsima{$\; \buildrel > \over \sim \;$}
\def\gsim{\lower.5ex\hbox{\gtsima}}

\title[On the July 2007 flare of the blazar 3C 454.3] 
{On the July 2007 flare of the blazar 3C 454.3}

\author[Ghisellini, Foschini, Tavecchio \& Pian] 
{Gabriele Ghisellini$^1$\thanks{E--mail:
gabriele.ghisellini@brera.inaf.it}, Luigi Foschini$^2$, Fabrizio Tavecchio$^1$ and Elena Pian$^3$\\
$^1$ INAF/Osservatorio Astronomico di Brera, via E. Bianchi 46, I--23807 Merate, Italy \\
$^2$ INAF/IASF--Bologna, Via Gobetti 101, I--40129 Bologna, Italy \\
$^3$ INAF/Osservatorio Astronomico di Trieste, via G.B. Tiepolo 11, I--34131, Italy \\
}

\begin{document}


\pagerange{\pageref{firstpage}--\pageref{lastpage}} \pubyear{2007}

\maketitle

\label{firstpage}

\begin{abstract}
In July 2007, the blazar 3C 454.3 underwent a flare in the optical,
reaching $R\sim 13$ on July 19.
Then the optical flux decreased by one magnitude, being $R\sim 14$ 
when the source was detected by the $\gamma$--ray satellite {\it AGILE},
that observed the source on July 24--30.
At the same time, the {\it Swift} satellite performed a series of snapshots.
We can construct the simultaneous spectral energy distribution 
using optical, UV, X--ray and $\gamma$--ray data.
These shows that an increased $\gamma$--ray flux is accompanied
by a weaker optical/X--ray flux with respect to the flare
observed in the Spring 2005 by {\it INTEGRAL} and {\it Swift}.
This confirms earlier suggestions about the behaviour of the jet of 3C 454.3.
\end{abstract}
\begin{keywords}
galaxies: active--galaxies: jets--galaxies: individual: 3C454.3 ---
radiation mechanisms: non--thermal
\end{keywords}

\section{Introduction}

Among the flat--spectrum radio quasar (FSRQ) class of blazars, 
3C 454.3 ($z=0.859$) is one of the brightest sources. 
It has been detected significantly at almost all the available wavelength, from radio 
to $\gamma$--rays (e.g., radio: Lister \& Homan 2005, Homan \& Lister 2006; infrared: 
Bennett et al. 2003; optical: Villata et al. 2006, Fuhrmann et al. 2006, 
ultraviolet: Pian et al. 2005, X--rays: Tavecchio et al. 2002, Pian et al. 2006, 
Giommi et al. 2006; $\gamma$--rays: Hartman et al. 1999, Zhang et al. 2005, 
Nandikotkur et al. 2007).
Its spectral energy distribution (SED) has the typical double--humped shape of blazars. 
In this case, the synchrotron peak is at 
infrared frequencies, while the inverse--Compton emission reach the maximum at MeV-GeV 
energies (Fossati et al. 1998, Ghisellini et al. 1998). 

In April--May 2005, 3C 454.3 literally exploded and had its historical 
maximum at optical wavelengths with $R$=12.0 (Villata et al. 2006). 
This exceptional event triggered observations at X--ray energies with \emph{INTEGRAL} 
(Pian et al. 2006) and \emph{Swift} (Giommi et al. 2006).
These detected the source up to 200 keV and in a brightened high energy state
(Pian et al. 2006). 
This strong optical and hard X--ray activity was followed by a radio outburst 
with about one year delay (Villata et al. 2007).

After a period of relative quiescence, the blazar underwent again an optical outburst 
in July 2007, although preliminary optical
observations\footnote{{\tt http://users.utu.fi/$\sim$kani/1m/index.html}} suggest a peak 
lower than that of 2005. 
The alert quickly spread all around the 
world\footnote{See the News 44 of the Whole Earth Blazar Telescope (WEBT) at 
\texttt{http://www.to.astro.it/blazars/webt/news.html}} and triggered observations 
with \emph{Swift} and \emph{AGILE} 
(Tavani et al. 2006).
Although the latter is still performing its verification phase, 
it was able to repoint towards the blazar on July 24--30, 2007 and 
to detect it (Bulgarelli et al. 2007).
The flux measured by \emph{AGILE} during the first part of monitoring, 
$F=(3\pm 1)\times 10^{-6}$ ph cm$^{-2}$ s$^{-1}$ at $E>$100 MeV (Vercellone et al. 2007), 
is a bit less than 3 times 
the maximum [$(1.25 \pm 0.19)\times 10^{-6}$~ph~cm$^{-2}$~s$^{-1}$]
and about 6 times greater than the average 
[$(5.6\pm 0.4)\times 10^{-7}$~ph~cm$^{-2}$~s$^{-1}$]
flux observed by EGRET (Nandikotkur et al. 2007). 

Here we report about the analysis and assembling of the simultaneous optical to 
$\gamma$--rays SED of 3C~$454.3$ in order to model this SED and test new
ideas about how to obtain dramatic flux changes in specific bands
without changing much the power carried by the jet, as detailed in 
the next section.

We use $H_0=70$ km s$^{-1}$ Mpc$^{-1}$  and $\Omega_{\rm M}=0.3$, 
$\Omega_{\Lambda}=0.7$.

\section{Scientific motivations}

During the huge multiwavelength flare observed in Spring 2005 
the optical flux increased by almost 
two orders of magnitude, while the X--ray flux
increased by a factor $\sim$10 with respect to the 2000 level
detected by {\it Beppo}SAX (Tavecchio et al. 2000).  
We can ask if this behavior corresponds to similar changes in the power
carried by the jet or if, instead, the jet power is nearly constant,
yet able to produce dramatic changes in specific energy bands.  
Pian et al. (2006) could indeed describe the observed SEDs in 
2000\footnote{
The 2000 SED actually includes also data taken at previous epochs. We
will hereafter refer to it as ``2000 SED" because the modelling of the SED 
is critically driven by the {\it Beppo}SAX spectrum, acquired in 2000.
}
and 2005 with minimal changes in the bolometric luminosity (and in the jet
power), assuming that the dissipation region (where most of the
radiation is produced) was inside the Broad Line Region (BLR) in 2000,
and outside it in Spring 2005.  If the magnetic field is the same in
the two epochs, despite the difference in the size of the emitting
regions, then the power output is dominated by the Inverse Compton
scattering of the broad line photons in 2000
(External Compton, EC), and by synchrotron and
synchrotron Self Compton (SSC) emission in 2005.

Then Katarzynski \& Ghisellini (2007) proposed an alternative model,
in which the location of the dissipation region varies along the jet,
and there is a link between this location, the compactness of the
source, and the value of the bulk Lorentz factor $\Gamma$.  The
smaller this distance, the smaller $\Gamma$, the larger the magnetic
field of the source.  The physical bases of this model were i) the
realization that the power lost in radiation must be a small fraction
of the jet power, which has to carry most of it to the radio lobe
regions (at least for powerful blazars); therefore ii) the kinetic and
Poynting power of the jet are likely conserved, implying that the
magnetic field $B\propto (\Gamma R)^{-1}$, where $R$ is the
(transverse) size of the jet; iii) in the internal shock model for
blazars the dissipation is due to the collision of two shells moving
with different $\Gamma$: the collision occurs at a distance which is a
factor $\Gamma^2$ larger than the initial shell separation.  This
introduces a link between compactness of the source and $\Gamma$; iv)
even if the dissipation is not due to internal shocks, we may have a
similar relation between $\Gamma$ and compactness if the jet is still
accelerating when dissipation occurs.  Following these ideas,
Katarzynski \& Ghisellini (2007) could reproduce the 2000 and 2005
states of 3C 454.3 with a jet of exactly the same power, by assuming
different $\Gamma$ (11 in 2000 and 6.25 in Spring 2005).  
In this model the observed bolometric luminosity is nearly constant (except for extreme
changes in $\Gamma$, resulting in extreme changes of the Doppler
factor and thus in the Doppler boosting of the observed radiation).
When dissipation occurs with large $\Gamma$ the seed external photons
are seen largely boosted, while the $B$--field is relatively weaker,
resulting in a EC dominated spectrum, where most of the power is
emitted the MeV--GeV band.  On the other hand, for smaller $\Gamma$, the
$B$--field is larger, the external radiation is seen less boosted,
resulting in a synchrotron and SSC dominated spectrum, with much less
contribution of the EC.

It is therefore clear that the main diagnostics of this model is the
level of the MeV--GeV flux during flares: if both the optical
(synchrotron) emission {\it and} the MeV--GeV (EC) emission increase,
then this flags a real increase of the jet power, but if the increase
of the MeV--GeV flux corresponds to the same (or even to a fainter)
optical flux, then this flags an emission episode by a jet of nearly
constant power but larger $\Gamma$.  Vice--versa, if the MeV--GeV
radiation decreases during an optical flare, this flags a jet of
constant power dissipating closer to the black hole and with a smaller
$\Gamma$.

We hope that these simple trends will be verified in the era of
simultaneous {\it GLAST} and {\it Swift} observations, able to 
look for the crucial part of the SED (optical to GeV bands).
But anticipations of this, for the brightest blazars,
are possible using {\it AGILE} and {\it Swift} observations.
This is exactly what happened for 3C 454.3 in July 2007.

\section{Data analysis of the {\it Swift} observations}

Although Swift observed 3C 454.3 several times in July and
August 2007, we analysed data from the only observation simultaneous with
\emph{AGILE}, i.e. that performed on July 26. 
The data from the three instruments onboard \emph{Swift} 
(Gehrels et al. 2004) have been processed and analysed with 
\texttt{HEASoft v. 6.3.1} with the CALDB release of July 30, 2007.  

Data from the X--ray Telescope XRT ($0.3-10$~keV, Burrows et al. 2005) 
were analysed using the \texttt{xrtpipeline} task,
set for the photon counting mode and having selected single to quadruple 
pixel events (grades $0-12$). The output data were rebinned in order to have 
at least 30 counts per energy bin. 
The final spectrum is best fitted ($\chi^2=45.9$ 
for 43 degrees of freedom) with a power law with absorption 
in excess of the Galactic value ($N^{\rm Gal}_{\rm H}=7.24\times 10^{20}$~cm$^{-2}$, 
Kalberla et al. 2005).
Formally, this additional $N_{\rm H}^{z}$ at $z=0.859$ 
has a value of $1.9_{-1.3}^{+1.5}\times 10^{21}$~cm$^{-2}$. 
The power law has a photon index of $1.47\pm 0.11$ and a 
flux of $6.1\times 10^{-11}$~erg~cm$^{-2}$~s$^{-1}$ in the $2-10$ 
keV energy band.

The additional $N_{\rm H}^{z}$ may be not physical, 
but it is used to fit satisfactorily the observed deficit of photons at low energies. 
A fit with a broken power--law model gives multiple minima for the low energy 
photon index, making it impossible to constrain efficiently the parameters. 
The photon deficit at low energies has a gradual decrease, resembling 
a low--energy cutoff, and the additional absorption is the best statistical 
way to mimic it. 
Therefore, when in the following we refer to the 
unabsorbed X--ray spectrum, we mean the spectrum corrected 
\emph{only} for the Galactic absorption.

Data from the UV telescope UVOT (Roming et al. 2005) were analysed 
by using the \texttt{uvotmaghist} task with a source region of $10''$, 
while the background was extracted from a source--free circular region 
with radius equal to $60''$ (it was not possible to use an annular region, 
because of a nearby source).  
We added a $10\%$ error in flux (corresponding to about 0.1 magnitudes) 
to take into account systematic effects. 
The observed magnitudes are $UVW1 = 14.7\pm 0.1$, 
$UVM2 = 14.9\pm 0.1$, and $UVW2=15.1\pm 0.1$.

Instead, for the hard X--ray detector 
BAT (optimised for the $15-150$ keV energy band, Barthelmy et al. 2005), 
it is necessary to perform long exposure observations in order to 
get useful signal from extragalactic sources. Therefore, data from 
all the July observations (performed on days 11, 12, 13, 14, 17, 20, 26, 28, 30) 
were binned, cleaned from hot pixels and background, and deconvolved. 
The intensity images were then integrated by using the variance as 
weighting factor and the total exposure was 15 ks. 
The flux  upper limits for a $3\sigma$ detection 
-- already corrected for systematics -- measured at the position of 
3C 454.3 are $2.1\times 10^{-10}$~erg~cm$^{-2}$~s$^{-1}$ (28 mCrab) 
and $3.6\times 10^{-10}$~erg~cm$^{-2}$~s$^{-1}$ (38 mCrab) in 
the 20--40 keV and 40--100 keV energy bands, respectively.

\begin{figure}
\hskip -2.3 true cm 
\psfig{file=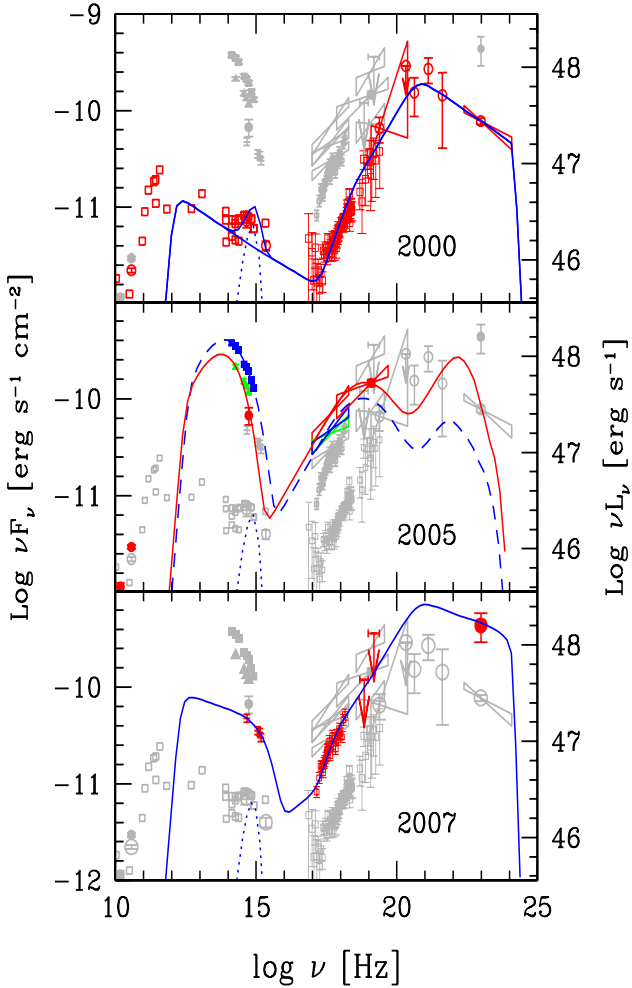,height=14cm,width=11.5cm}
\vskip -0.5 true cm
\caption{SEDs of 3C 454.3 at different epochs. 
Top panel: the SED in 2000, corresponding to the {\it Beppo}SAX
observations discussed in Tavecchio et al. (2002).
The other points are not simultaneous. 
(see Pian et al. 2006 and references therein).
Mid panel: the SED during the huge optical flare in 2005,
as described in Pian et al (2006) and Giommi et al. (2006).
Different colors correspond to different sets of strictly
simultaneous data.
Bottom panel: the SED on July 26, 2007, as observed by {\it AGILE},
and {\it Swift}. We have also added the optical flux in the $R$
band, taking the average between the magnitude of July 24 and July 29
measure by the Tuorla observatory.
The optical and X--ray fluxes are corrected for Galactic extinction
($A_V=0.355$).
The solid and dashed lines correspond to our modelling (parameters in Tab. \ref{para}).
The dotted line is the contribution from the accretion disk (assumed
to be a simple black--body). 
}
\label{agile}
\end{figure}

\section{Modelling the SED}

We adopt a simple, one--zone, homogeneous synchrotron and 
inverse Compton model to the different states of 3C 454.3.
The model is fully described in Ghisellini, Celotti \& Costamante
(2002), and we briefly outline here the general assumptions.

The source is a cylinder of cross sectional radius
$R=\psi z$, where $z$ is the distance from the apex of the jet,
assumed to be a cone of semiaperture angle $\psi\sim 0.1$.
The width $\Delta R^\prime$, as measured in the comoving frame, is
assumed to be within the range $R/\Gamma \le \Delta R^\prime \le R$.
The magnetic field $B$ is homogeneous and tangled;  
the blob moves with a bulk Lorentz factor $\Gamma$ 
and the viewing angle is $\theta$.
The external radiation is produced at a fixed radius, that can 
be identified with the radius $R_{\rm BLR}$ of the broad line 
region (BLR; Sikora, Begelman \& Rees 1994), reprocessing 
$\sim$10\% of the disk luminosity.
This is likely an approximation of the real amount of external 
radiation, since, besides the contribution of the BLR, other 
processes can contribute (scattering by 
ionised intercloud plasma, synchrotron radiation ``mirrored'' by 
the clouds and/or the walls of the jet, Ghisellini \& Madau 1996, 
and also direct radiation from the accretion disk, 
Dermer \& Schlickeiser 1993; Celotti, Ghisellini \& Fabian, 2007).

The particle distribution is the result of injection (for a finite time,
taken to be equal to $\Delta R^\prime/c$ in the comoving frame) and cooling.
We calculate the random Lorentz factor $\gamma_{\rm cool}$ at which
the particles cool in one light crossing time.
If the particles are injected between $\gamma_{\rm inj}$ and $\gamma_{\rm max}$
with a power law distribution of slope $s$, 
in the ``fast cooling'' regime
($\gamma_{\rm cool}<\gamma_{\rm inj}$), we have
$N(\gamma) \propto \gamma^{-p}$ between $\gamma_{\rm inj}$ and $\gamma_{\rm max}$,
(where $p=s+1$ is the injection slope increased by one unit)
and 
$N(\gamma) \propto \gamma^{-2}$ between $\gamma_{\rm cool}$ and $\gamma_{\rm inj}$.
In the models presented here, this is always the case, since $\gamma_{\rm cool}$
is of order unity in all cases.

Based on the above assumptions, the SED modelling yields the results shown
in Fig. \ref{agile}, where we report model curves for four states of the source.
The upper panel shows the SED corresponding to the {\it Beppo}SAX 
observations in 2000, with which we associate the (not simultaneous) 
{\it COMPTEL}, {\it EGRET} and optical data 
(Tavecchio et al. 2002; Pian et al. 2006; Hartman et al. 1999; Zhang et al. 2005). 
The mid panel shows two states in 2005, presented in Pian et al. (2006) and
Giommi et al. (2006),
for which we have simultaneous optical, soft and hard X--ray {\it INTEGRAL} data.
The bottom panel shows the state in July 26, 2007, as observed
by \emph{Swift} and \emph{AGILE}, together with an optical point (in the $R$ band)
taken during the multiwavelength campaign of this source by the
Tuorla observatory.

The input parameters used for these models are listed in 
the upper part of Tab. \ref{para}. 
In the lower part we report the derived
Doppler factor $\delta=1/[\Gamma (1-\beta\cos\theta)]$ and the power carried 
by the jet in the form of magnetic field ($L_B$), cold protons
($L_{\rm p})$, relativistic electrons ($L_{\rm e}$), 
and produced radiation ($L_{\rm rad}$), defined as:
\begin{eqnarray}
L_B\,         &=& \,  \pi R^2 \Gamma^2 c \, U_B
\nonumber \\ 
L_{\rm p}\,   &=& \,  \pi R^2 \Gamma^2 c \int N(\gamma) m_{\rm p} c^2 d\gamma
\nonumber \\
L_{\rm e} \,  &=& \,  \pi R^2 \Gamma^2 c \int N(\gamma) \gamma m_{\rm e} c^2 d\gamma
\nonumber \\
L_{\rm rad} \, &=& \, \pi R^2 \Gamma^2 c \, U^\prime_{\rm rad} 
\end{eqnarray}
where $U_B$ and $U^\prime_{\rm rad}$ are the magnetic and
the radiation energy density measured in the comoving frame, respectively,
and where we have assumed one proton per relativistic electron.

The choice of the parameters is not unique in SSC+EC models,
especially when the high energy ($\gamma$--ray) peak
is not observed, as in 2005.
The adopted choice reflects the following considerations:
i) we have used the low optical state in 2000 (top panel) to
fix the contribution of the accretion disk luminosity;
ii) we have tried to minimise the total power budget;
iii) we have tried to describe the different states of the source
with a minimal change of the power carried by the jet.

We find that the July 2007 flare and the state in 2000
can be reproduced by very similar parameters
(same radius and bulk Lorentz factors, slightly different
magnetic field and shape of the particle distribution),
the main difference being the amount of injected power
($L_{\rm inj}^\prime$ going from $7\times 10^{43}$ to
$3\times 10^{44}$ erg s$^{-1})$.
For the two states in 2005, instead, the SED can be reproduced
by a more compact source, with a smaller $\Gamma$, corresponding
to a smaller Doppler factor $\delta$.
The magnetic field is greater ($B=15$--35 G), and the injected power 
is slightly greater than for the 2007 state.

The model can reproduce the huge variations observed
especially in the optical (between 2000 and 2005 there is a difference
of a factor $\sim$100) and in the $\gamma$--ray band (the 2007 state
is one order of magnitude larger than in 2000).  This is achieved with
much smaller variations of the power transported by the jet, as
discussed below.

Note that in all cases the emission drops above 50--100 GeV. 
This is due to the robust limit to the maximum emitted $\gamma$--ray energy 
provided by the decrease of the Klein--Nishina cross section.
The maximum energy can be estimated to be
$h\nu_{\rm max}\simeq 63/ \nu_{t,15}$ GeV, where $\nu _{t,15}$ is the
frequency of the target soft photons (as seen in the observer frame)
in units of $10^{15}$ Hz. 
Internal (i.e. inside the source) photon--photon absorption, instead, plays no role 
in the models presented here.

\begin{table}
\begin{center}
\caption{For all models we have assumed a viewing angle $\theta=3.5^\circ$ 
and a bolometric luminosity of the accretion disk 
$L_{\rm disk}=3\times 10^{46}$ erg s$^{-1}$. 
We assume that the 10\% of the disk luminosity
is reprocessed by the BLR and re--emitted as broad lines
at a distance $R_{\rm BLR}=3\times 10^{17}$ cm for all models.
The minimum variability timescale $t_{\rm var}$ is defined
as $t_{\rm var}=R/(c\delta)$.
}
\label{para}
\begin{tabular}{|l|lllll|}
\hline
\hline
                     &2000  &2005 blue &2005 red &2007  &Units \\
\hline
$\Gamma$             &16    &7         &8        &16     &       \\
$R$                  &25    &15        &17       &25     & [$10^{15}$ cm]   \\
$\Delta R^\prime$    &10    &5         &9        &13     & [$10^{15}$ cm]   \\
$B$                  &7     &35        &15       &9      &  [G]    \\
$\gamma_{\rm peak}$  &50    &300       &400      &50     &     \\
$\gamma_{\rm max}$   &40    &1         &1        &2.7    &  [$10^{3}$]   \\
$p$                  &3.4   &4.0       &4.0      &3.2    &     \\
$L^\prime_{\rm inj}$ &0.07  &0.5       &0.4      &0.3    & [$10^{45}$ erg s$^{-1}$] \\
\hline
$\delta$             &16.4  &11.8      &12.9     &16.4   &              \\
$t_{\rm var}$        &14    &11.8      &12.2     &14     &hours         \\
$L_{B}$              &29.3  &50.1      &15.5     &48.5   &[$10^{45}$ erg s$^{-1}$]  \\
$L_{\rm p}$          &444.4 &180       &227      &1549   &[$10^{45}$ erg s$^{-1}$]  \\
$L_{\rm e}$          &1.55  &0.75      &1.13     &4.7    &[$10^{45}$ erg s$^{-1}$]  \\
$L_{\rm rad}$        &17.4  &22.2      &24.8     &77.5   &[$10^{45}$ erg s$^{-1}$]   \\
\hline
\end{tabular}
\end{center}
\end{table}

\begin{figure}
\vskip -0.5 true cm
\hskip -0.1 true cm
\psfig{file=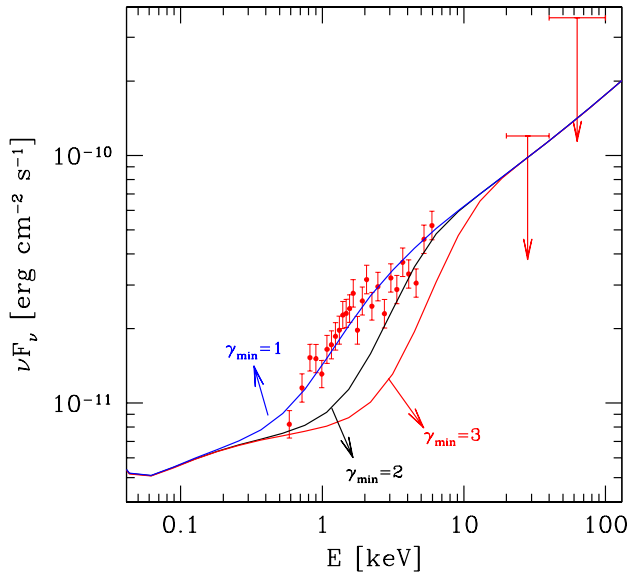,height=9.cm,width=9.cm}
\caption{
X--ray data of July 2007 with, superimposed,
the result of our model assuming that the minimum energy of the
emitting electrons corresponds to $\gamma_{\rm min}=1$, 2 and 3,
as labelled.
Only for $\gamma_{\rm min}=1$ we obtain a reasonable fit. 
}
\label{gmin}
\end{figure}

\begin{figure}
\psfig{file=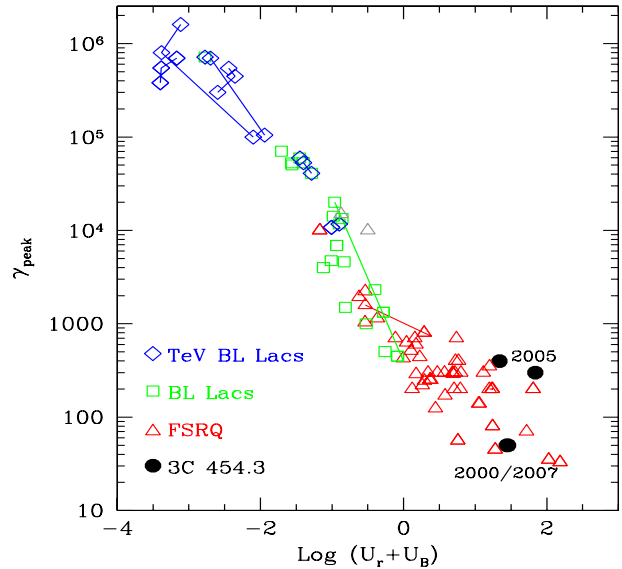,height=8.cm,width=9.5cm}
\caption{The location of the different states of 3C 454.3 
in the $\gamma_{\rm peak}$--$(U_{\rm r}+U_{\rm B})$ plane.
Adapted from Foschini et al. (2006).
}
\label{gbu}
\end{figure}

\section{Discussion}

Our modelling of the SED of 3C 454.3 in 2000, 2005 and 2007 
is in qualitative agreement with the expectations of the jet 
radiation model proposed by Katarzynski \& Ghisellini (2007), 
where the relative importance of the synchrotron and SSC luminosity 
(with respect to the EC one) is controlled by the value of $\Gamma$, 
which is associated with the compactness of the source.
The relation between $\Gamma$ and source compactness relies either
on the assumption that the jet is still accelerating when it dissipates, or on
the assumption that the dissipation is the result of internal shocks between 
two shells (Spada et al. 2001), 
occurring at a distance $\sim \Gamma^2$ times the initial separation: 
in both cases we have that emitting regions with smaller $\Gamma$ 
are closer to the black hole, and thus more compact.
The magnetic field would be stronger, implying larger
synchrotron and SSC luminosities with respect to the EC one.
It is then possible to have dramatic variations in specific frequency
bands even for jets carrying the same amount of power.

Indeed, the power in bulk motion of the relativistic electrons
(see Tab. \ref{para}) varies no more than a factor 6.3 
for the 4 different states we have considered,
while the Poynting flux varies merely by a factor 3.
The largest change of the jet power is due to the
proton component (factor 8.6), assuming that there is,
for all states, one proton per relativistic electron, and
that the electron distribution extends down to $\gamma_{\rm min}=1$.
The latter assumption is usually very difficult to test, since
low energy electrons emit synchrotron radiation in the
self--absorbed regime (which is both insensitive to the low--energy
cutoff and practically impossible to observe)
and SSC radiation at relatively low frequencies, where the flux
is dominated by the synchrotron emission of electrons of higher energies.
The best opportunity to see the radiation emitted by electrons of low 
$\gamma$ is thus when the SSC contribution is not contributing much to the 
emission in the X--ray band, which is instead dominated by EC radiation.
In fact in this band the EC is produced by low energy electrons.
This is the case for the 2000 and 2007 states of 3C 454.3.
Fig. \ref{gmin} shows the predicted spectrum varying $\gamma_{\rm min}$
from 1 to 3. 
It shows that even tiny variations have a large effect on the spectrum,
which becomes much harder than observed when $\gamma_{\rm min}> 1$.
We then conclude that the electron distribution {\it must}
extend down to $\gamma_{\rm min}\sim 1$.

Consider now the value of $L_{\rm rad}$ in comparison with the other powers.
It is of the order of, but in some case larger, than $L_{\rm e}$ and
$L_{\rm B}$.
This shows that the power spent by the jet to produce the radiation we see
cannot be provided by the magnetic field or by relativistic electrons only.
There must be another form of power, able to re--furnish the electrons
of the energy lost for producing the radiation.
The simplest possibility is that the jet is baryon loaded, and that $L_{\rm p}$
dominates the jet power.
In our case (assuming one proton per electron) we can estimate the
radiative efficiency (i.e. $L_{\rm rad}/L_{\rm p}$) which is between
4 and 12 per cent.
This is in excellent agreement with the idea that the jet {\it must}
be radiatively inefficient, since it has to carry most of its power to
the radio lobes 
(see also Celotti \& Ghisellini 2007).

Finally, we can see where the different states of 3C 454.3 
are located in the plane $\gamma_{\rm peak}$--$U$, where
$\gamma_{\rm peak}$ is the random Lorentz factor
of the electrons emitting at the two peaks of the SED,
and $U=U_B+U_{\rm rad}$ is the radiation+magnetic energy
density as seen in the comoving frame.

In this plane blazars form a well defined sequence
(see Ghisellini et al. 1998; 2002): low power BL Lacs
follow  a relation $\gamma_{\rm peak}\propto U^{-1}$
and more powerful blazars (FSRQ with broad emission lines)
tend to lie on a $\gamma_{\rm peak}\propto U^{-0.6}$ relation.
As Fig. \ref{gbu} shows, the 4 states of 3C 454.3 considered
here are somewhat scattered, but they are within the total
scatter of the relation in this high $U$, small $\gamma_{\rm peak}$ part.
Note also that in this plane the 2000 and 2007 states 
(of strong $\gamma$--ray flux) are identical.

We can conclude that the July 2007 flare of 3C 454.3 gives support
to the idea that dissipation site in relativistic jets changes,
and that it is likely that, when it is located closer to the
black hole, the emitting region is more compact, has a smaller
bulk Lorentz factor, and a greater magnetic field, although the
power in Poynting flux needs not to change, since a larger 
$B$--field is compensated by a smaller size and bulk Lorentz factor,
making $L_{\rm B}$ nearly constant.
Observationally, these characteristics translate in dramatic
variations of the synchrotron peak flux and of the EC high
energy component, and in an inverse correlation between
the synchrotron and EC components.
The more compact the source is, the more intense
its synchrotron flux, at the expenses of the EC emission.
The latter becomes fainter for the weaker Doppler boosting of 
the external seed photons in the comoving frame, due to the smaller 
bulk Lorentz factor.
The flux variations at high energies are however mitigated by the
SSC emission, which is enhanced for more compact emission regions.
This component can dominate in the X--ray band, and can be recognized
by a softer and more intense X--ray spectrum with respect to the spectrum
due to the EC process.
Since in this case the production of high energy $\gamma$--rays is
reduced, the pair production process, due to photon--photon interactions, 
is reduced also, and we can evade the problem of the reprocessing of 
high energy $\gamma$--ray luminosity into the X--ray band, which 
inhibits the production of GeV radiation in too compact sources.
If different states are indeed related to different compactness of the
source, one might also expect different variability timescales.
However (and unfortunately) this is mitigated by the different 
Doppler factors, which are smaller for more compact emitting regions,
resulting (see Table 1) in minimum variability timescales
$t_{\rm var}=R/(c\delta)$ that are almost equal.

\section*{Acknowledgements}
We thank the Tuorla Observatory AGN optical monitoring team, Finland, 
for giving us the permission to use their $R$--band data.
We acknowledge the use of public data from the Swift data archive.
This research has made use of data obtained from the High 
Energy Astrophysics Science Archive Research Center (HEASARC), 
provided by NASA's Goddard Space Flight Center.
Support from the Italian Space Agency, under contract ASI-INAF
I/023/05/0 is acknowledged.

\end{document}